\documentclass[a4paper,11pt]{article}
\usepackage{pos}

\usepackage{amsfonts} 
\usepackage{amssymb} 
\usepackage{amsmath} 
\usepackage{graphicx} 
\usepackage{subfigure} 
\usepackage{array} 
\usepackage{dcolumn} 
\usepackage{bm} 
\usepackage{latexsym} 
\usepackage{longtable} 
\usepackage{hyperref} 
\usepackage{bm}
\usepackage{slashed}
\usepackage{bbold}
\usepackage{color}
\usepackage{multirow}
\usepackage{rotating}
\usepackage{comment}
\usepackage [english]{babel}
\usepackage [autostyle, english = american]{csquotes}
\MakeOuterQuote{"}
\usepackage{graphicx}
\usepackage{subfigure}
\usepackage{xcolor,cancel}
\usepackage{appendix}[toc,page]
\usepackage{setspace}
\usepackage{amsmath, amssymb}
\usepackage{slashed}

\newcommand{\be}{\begin{eqnarray}}
\newcommand{\ee}{\end{eqnarray}}

\newcommand{\CPV}{$\cancel{\text{CP}}$}

\providecommand{\matrixe}[3]{\langle#1\lvert#2\rvert#3\rangle}

\newcommand{\kk}{\mathbf{k}}

\newcommand{\spiN}{\sigma_{\pi N}}

\title{Excited states and precision results for nucleon charges and form factors}
\ShortTitle{Excited states and nucleon matrix elements}

\author*[a]{Rajan Gupta}
\author[a]{Tanmoy Bhattacharya}
\author[a]{Vincenzo Cirigliano}
\author[b]{Martin Hoferichter}
\author[c]{Yong-Chull Jang}
\author[d]{Balint Joo}
\author[a]{Emanuele Mereghetti}
\author[a,e]{Santanu Mondal}
\author[a,g]{Sungwoo Park}
\author[g]{Frank Winter}
\author[h]{Boram Yoon}

\affiliation[a]{Los Alamos National Laboratory, Theoretical Division T-2, Los Alamos, NM 87545, USA}

\affiliation[b]{Albert Einstein Center for Fundamental Physics, Institute for Theoretical Physics, University of Bern, Sidlerstrasse 5, 3012 Bern, Switzerland}
\affiliation[c]{Physics Department, Columbia University, New York, NY 10027, USA}
\affiliation[d]{Oak Ridge Leadership Computing Facility, Oak Ridge National Laboratory, Oak Ridge, TN 37831, USA}

\affiliation[e]{Department of Physics and Astronomy, Michigan State University, MI, 48824, USA}
\affiliation[f]{Department of Computational Mathematics, Science and Engineering, Michigan State University, MI, 48824, USA}

\affiliation[g]{Jefferson Lab, 12000 Jefferson Avenue, Newport News, Virginia 23606, USA}
\affiliation[h]{Los Alamos National Laboratory, Computer Computational and Statistical Sciences Division, CCS-7, Los Alamos, NM 87545, USA}

\emailAdd{rajan@lanl.gov}

\abstract{The exponentially falling signal-to-noise ratio in all
  nucleon correlation functions, and the presence of towers of
  multihadron excited states with relatively small mass gaps makes
  extraction of matrix elements of various operators within the ground
  state nucleon challenging. Theoretically, the allowed positive
  parity states with the smallest mass gaps are the $N(\bm p)\pi(-\bm
  p)$, $N(\bm 0)\pi(\bm 0)\pi(\bm 0)$, $N(\bm p)\pi(\bm 0)$, $N(\bm
  0)\pi(\bm p),\ \ldots$, states. A priori, the contribution of these
  states arises at one loop in chiral perturbation theory ($\chi$PT),
  however, in many cases the contributions are enhanced.  In this
  talk, I will review four such cases: the correlation functions from
  which the axial form factors, electric and magnetic form factors,
  the $\Theta$-term contribution to neutron electric dipole moment
  (nEDM), and the pion-nucleon sigma term are extracted. Including
  appropriate multihadron states in the analysis can lead to significantly
  different results compared to standard analyses with the mass gaps
  taken from fits to 2-point functions. The $\chi$PT case for $N \pi$ states is
  the most clear in the axial/pseudoscalar form factors which need to 
  satisfy the PCAC relation between them.  Our analyses, supported by $\chi$PT,
  suggests similarly large effects in the calculations of the
  $\Theta$-term and the pion-nucleon sigma term that have significant
  phenomenological implications. }

\FullConference{%
 The 38th International Symposium on Lattice Field Theory, LATTICE2021
  26th-30th July, 2021
  Zoom/Gather@Massachusetts Institute of Technology
}

\begin{document}
\maketitle
\section{Introduction}

Precision calculations of the matrix elements of various local and
nonlocal operators composed of quark and gluon fields provide detailed
knowledge of the hadron structure. Simulations are being done with
realistic values of the light (mostly assuming isospin symmetry, $m_u=
m_d$), strange and charm quark masses over a range of lattice spacings
$0.04 \lesssim a \lesssim 0.15$~fm that provide good control over
discretization
errors~\cite{Aoki:2019cca,Park:2021ypf,Aoki:2021kgd}. Finite volume
corrections in nucleon properties are observed to be small for $M_\pi
L \ge 4$~\cite{Aoki:2019cca,Park:2021ypf,Aoki:2021kgd}.  The main
challenges to obtaining percent level results are statistical errors, 
excited-states contributions (ESC), and the concomitant unresolved chiral
behavior.  With $O(10^6)$ measurements on about 5000 configurations,
one can get a good statistical signal up to $\approx 2$~fm in 2-point
correlation functions and up to $\approx 1.5$~fm in 3-point
functions~\cite{Park:2021ypf}. Even at these source-sink separations,
ESC are found to be large. To remove these contributions using fits to
the spectral decomposition of these correlation functions requires
knowing the energies of the excited states that make significant
contributions. Possible states include radial excitations and towers
of multihadron states, $N (\bm p) \pi(-\bm p)$, $N \pi \pi$, $\ldots$,
characterized by relative momenta $\bm p$ and having the quantum
numbers of the nucleon.  This talk discusses three quantities whose
values extracted using the standard analysis (using the spectrum
obtained from fits to the 2-point functions) that misses the $N\pi$
states differs very significantly from those obtained including the
lowest allowed $N \pi$, $N \pi \pi$ states that are motivated by
chiral perturbation theory ($\chi$PT).  Resolving all the excited
states that contribute significantly to a given correlation function
is, therefore, essential to progress.

\section{Spectrum from nucleon 2-point function}

The spectrum in a finite box of a nucleon with momentum $\bm p ={\bf n}2 \pi/La$ can be determined 
from fits to the spectral decomposition of the two-point function
$C^{2\text{pt}}$:
\begin{equation}
  C^{2\text{pt}}(\tau;\bm{p}) = \sum_{i} |{\mathcal{A}_i(\bm p)}|^2 e^{-E_i(\bm p) \tau}  \,.
\label{eq:2pt}
\end{equation}
Here $E_i$ are the energies and ${\mathcal{A}}_i$ are the
corresponding amplitudes for the creation/annihilation of a given
state $|i\rangle$. In all the calculations discussed here, 
%
$\mathcal{N}(x) = \epsilon^{abc} \left[ {q_1^a}^T(x) C \gamma_5  \frac{(1 \pm \gamma_4)}{2} q_2^b(x) \right] 
q_1^c(x) $
was used for the interpolating operator at both the source and the sink. In this setup, 
states with small ${\mathcal{A}}_i$ will be missed in the fit to Eq.~\eqref{eq:2pt} and thus in the 
standard analysis of 3-point functions.

\begin{figure}[htbp]
\centering
\includegraphics[angle=0,width=0.42\textwidth]{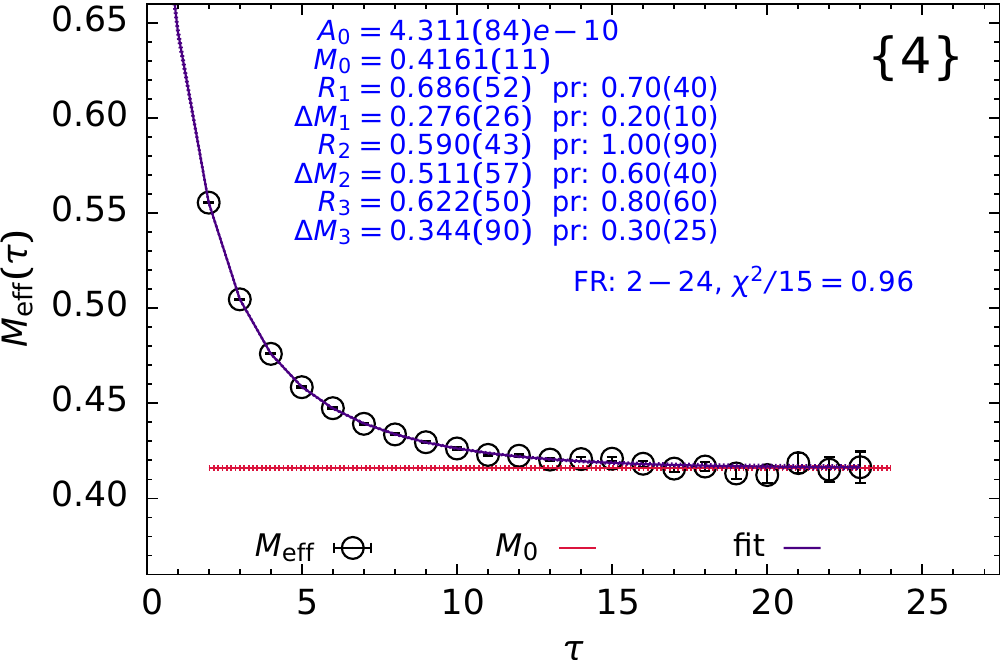}
\includegraphics[angle=0,width=0.42\textwidth]{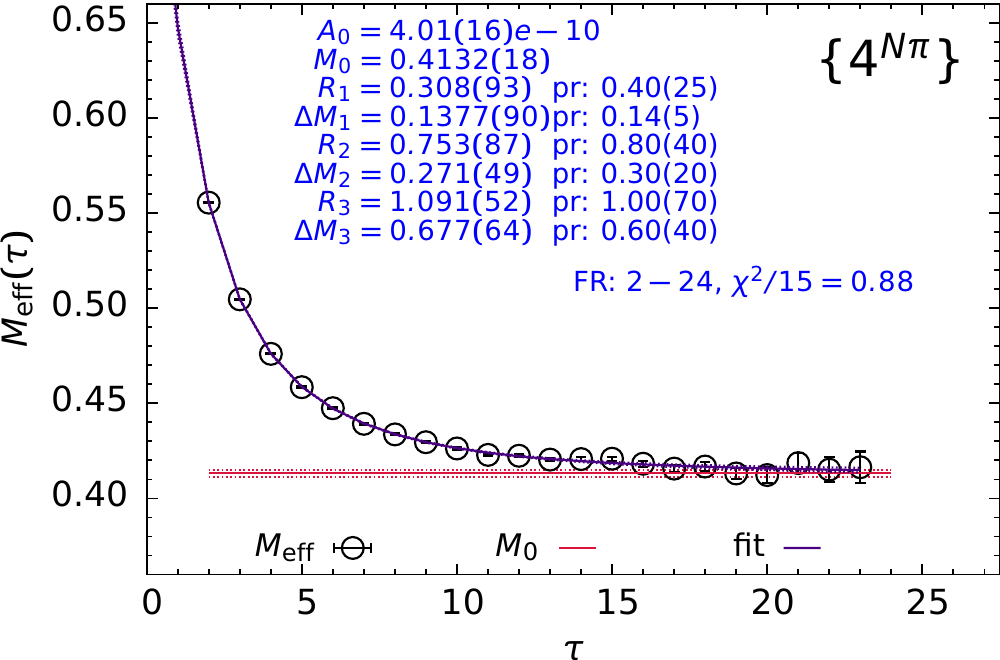}
\caption{Nucleon effective mass plot using data from $64^3 \times 128$
  lattices at $a=0.091$~fm and $M_\pi = 170$~MeV.  The statistical
  sample consists of 3000 configurations with 320 measurements on each
  configuration.  Errors on the data are calculated using a single
  elimination jackknife procedure on data binned over 6
  configurations. }
\label{fig:2pt}
\end{figure}

An example of the conumdrum of ESC is shown in Fig.~\ref{fig:2pt} using high
statistics data and fit to Eq.~\eqref{eq:2pt} truncated at four
states. (See Ref.~\cite{Park:2021ypf} for details.) The left panel
shows the standard analysis with wide priors used only to stabilize
the fit, while the right panel shows a fit with a narrow prior for
$E_1$ taken to be the energy of a non-interacting $N (\bm 1) \pi(-\bm
1)$ state. The resulting $E_1$ are about $ 1.5$ and 1.2~GeV,
respectively. The two outcomes are not distinguished by the augmented
$\chi^2$ minimized in the fits. In fact, in 4-state fits there is a
whole region of parameter space that gives similar $\chi^2$ in which
$1.2 < E_1 < 1.5$~GeV is equally likely.  Furthermore, assuming $R_1
\equiv |{\mathcal{A}}_1/{\mathcal{A}}_0|^2 =1$, the contribution of a
state with $\Delta E_1 = 300$~MeV is still $20\%$ (5\%) at $\tau/a =
11$ (22), i.e., at source-sink separation $\tau$ of 1fm (2fm). Thus, very
high precision data at $\tau/a > 1$~fm are needed to resolve
ES. \looseness-1

Bottom line: methods to beat the exponential, $e^{-(M_N -
  1.5M_\pi)\tau}$, decay in signal in all nucleon correlation
functions are needed.
Recognizing that the excited state spectrum is not easy to resolve
from fits to Eq.~\eqref{eq:2pt} with data generated using a single
interpolating operator $\cal N$, developing the computationally
challenging technology for using a variational basis of operators
including multihadron states such as $N \pi$, $N \pi \pi$ is
essential.
Motivation for including $N \pi$, $N \pi \pi$,
$\dots$ states in the analysis of 3-point functions comes from 
theory ($\chi$PT), and in the case of the axial form
factors from satisfying the PCAC relation as discussed
next. Throughout this talk, ``standard'' analysis will imply using the
spectrum from fits to the 2-point correlator ($E_1 \geq $ 
$N(1440)$), whereas ``$N \pi$'' analysis implies that $E_1$ used is
essentially the non-interacting energy of the $N \pi$ state calculated
on that ensemble.\looseness-1

\vspace{-0.2in}
\begin{figure}[htbp]
\centering
\includegraphics[angle=0,width=0.32\textwidth]{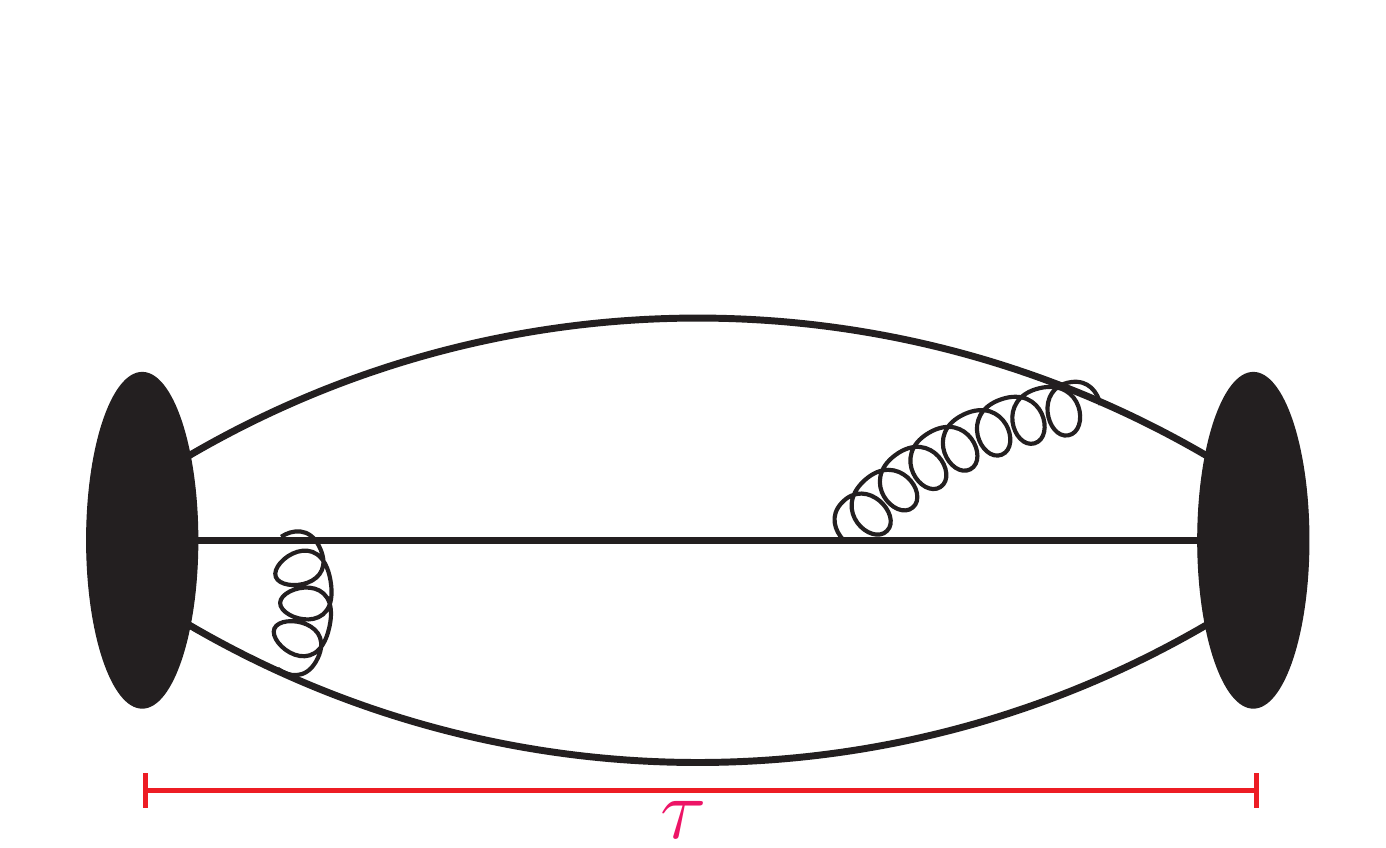}
\includegraphics[angle=0,width=0.32\textwidth]{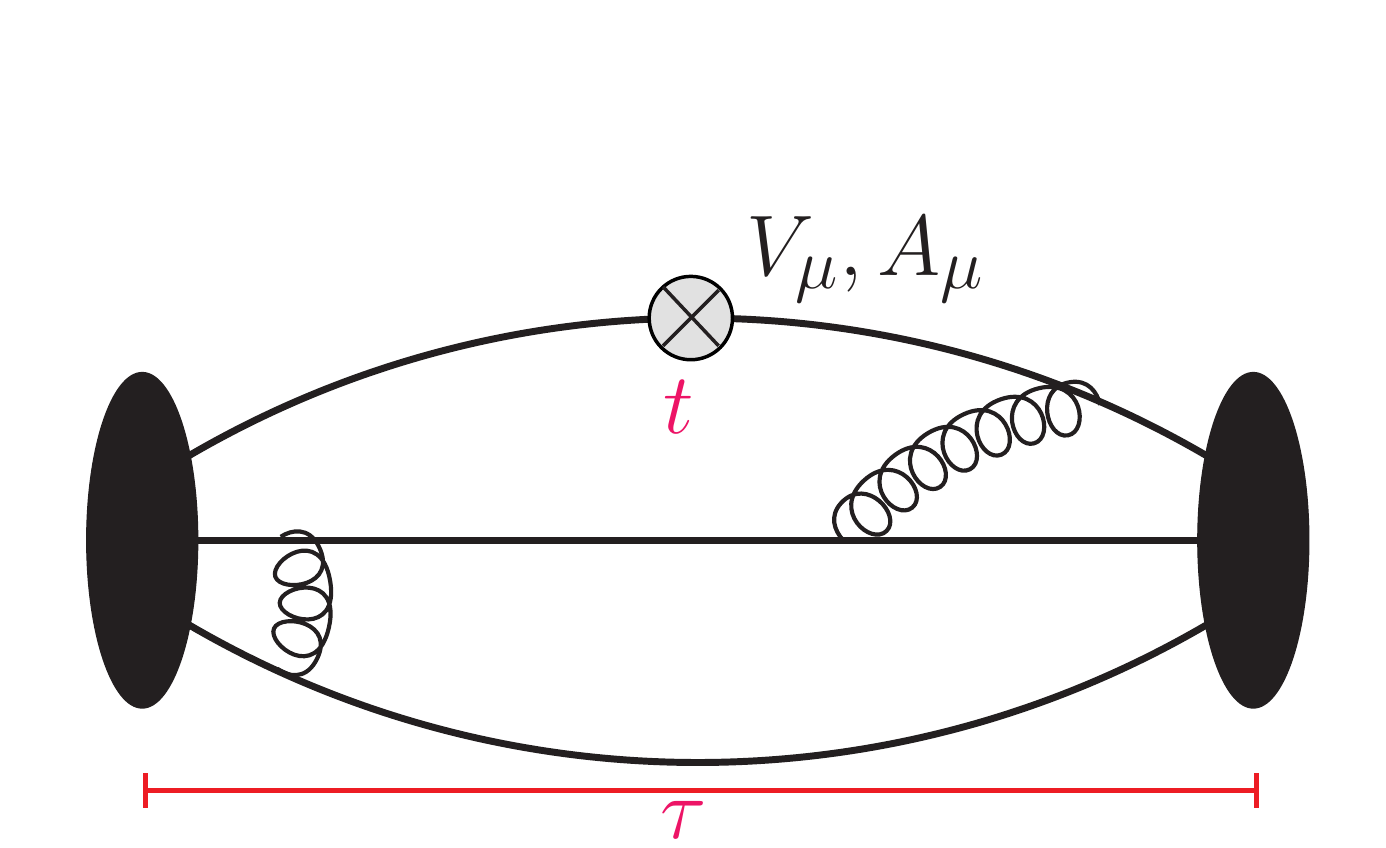}
\includegraphics[angle=0,width=0.32\textwidth]{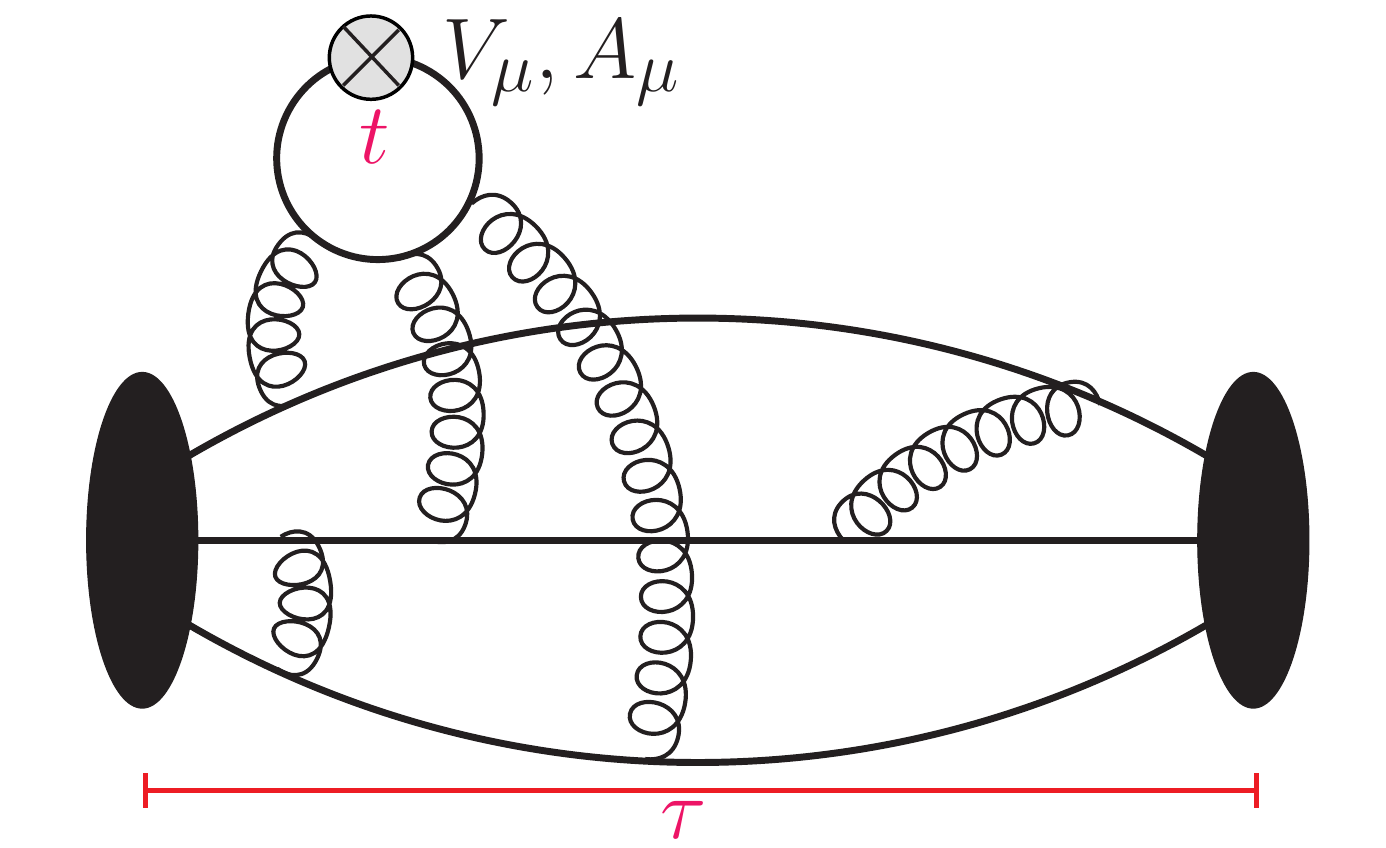}
\caption{Quark line diagrams illustrating the 2-point (left), 3-point
  connected (middle) and 3-point disconnected (right) correlation
  functions with the insertion of the vector $V_\mu$ or the axial
  $A_\mu$ currents. }
\label{fig:3pt}
\end{figure}

\section{Excited states in 3-point functions}

The spectral decomposition of any three-point
function $C_\mathcal{O}^{3\text{pt}}$ (see Fig.~\ref{fig:3pt} for illustration) is:
\begin{equation}
  C_\mathcal{O}^{3\text{pt}}(\tau;t) =
   \sum_{i,j} {\mathcal{A}_i} {\mathcal{A}_j^{-\bm q}}\matrixe{i}{\mathcal{O}_{\bm p}}{j^{-\bm q}} e^{-E_j t - E_i(\tau-t)}\,. 
   \label{eq:3pt}
\end{equation}
The operator inserts momentum $\bm q$, $\mathcal{N}$ at the sink is
projected to $\bm p =0$, making the nucleon momentum ${\bm p} = -\bm
q$ at the source. The spin projection operator used for the forward
propagating nucleon is $(1+\gamma_4)(1+i\gamma_5\gamma_3)/2$. It is
important to note that for $\bm q \neq 0$ case relevant to form
factors, the allowed excited states on the two sides of the operator
are different: for example they can be $N(\bm k)\pi(-\bm k), \forall
{\bm k}\neq 0$ on the ${\bm p} = 0$ side and $N(\bm k)\pi(-\bm {(p
  +k)})$ or $N(-\bm{(p+k})\pi(\bm k), \forall {\bm k}$ on the ${\bm p}
\neq 0$ side. Thus different towers of multihadron, in addition to
single-particle, excited states contribute.\looseness-1

At 1-loop in $\chi$PT, there is a long distance pion loop in all
possible configurations~\cite{Bar:2018xyi} which can, a priori, give a
large correction to any 3-point function. The question is whether this
or higher order contributions are significant? We discuss three cases
where these ESC are, in fact, enhanced, and one case (EM form factors)
where they are not.

\begin{figure*}[tbp] 
\subfigure
{
    \includegraphics[width=0.30\linewidth]{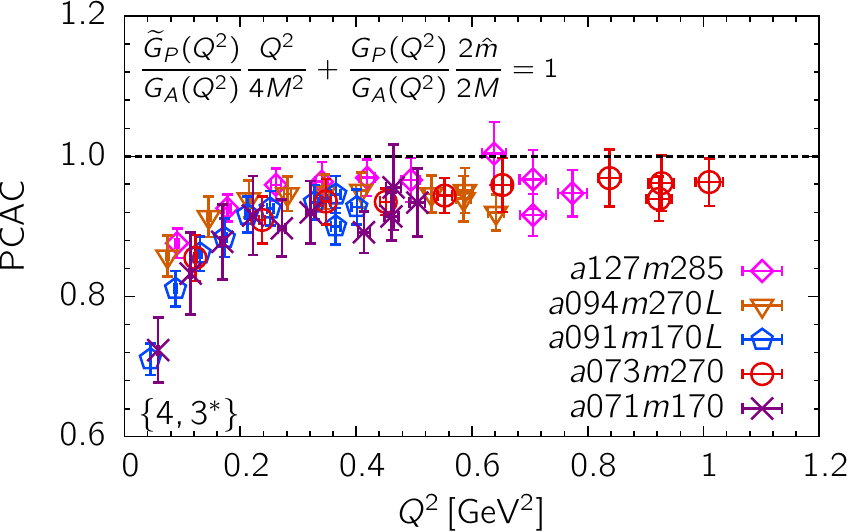}  
    \includegraphics[width=0.30\linewidth]{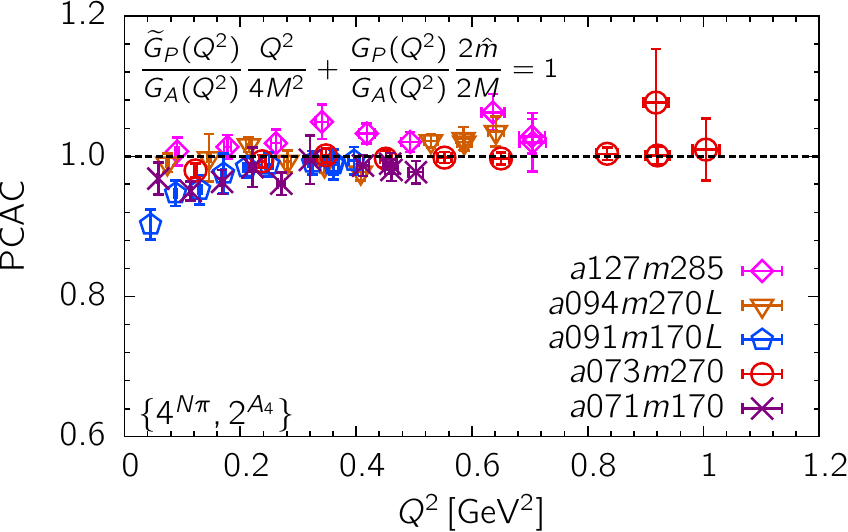}  
    \includegraphics[width=0.30\linewidth]{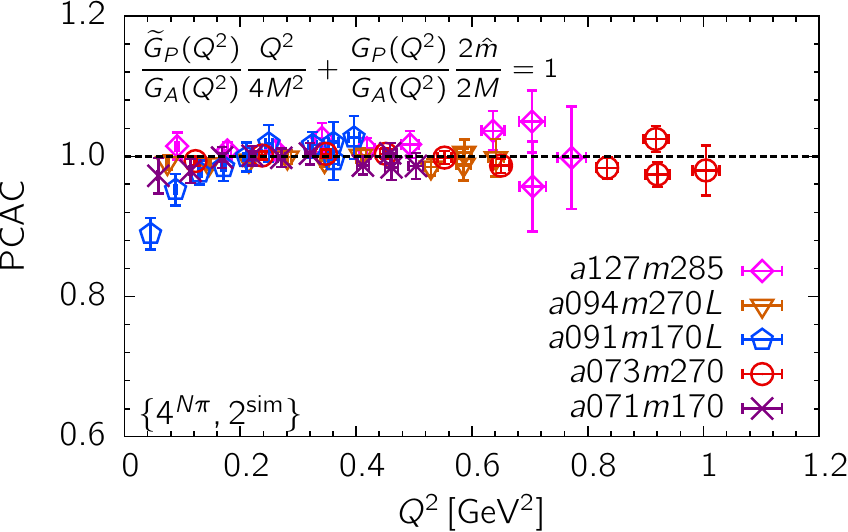}  
}
\vspace{-0.1cm}
\caption{The three panels show the degree to which the axial form
  factors satisfy the PCAC relation, Eq.~\protect\eqref{eq:PCAC}, for
  three analysis strategies specified at the bottom left corner and
  described in the text. Data from Ref.~\protect\cite{Park:2021ypf}.
\label{fig:PCAC}}
\end{figure*}
\vspace{-0.1cm}

\section{Axial vector form factors}

The axial and pseudoscalar form factors, $ G_A(q^2),
{\widetilde{G}_P(q^2)}\ {\rm and}\ G_P(q^2)$ are obtained by
decomposing the following matrix elements (ME) calculated within the
nucleon ground state $| N({\bm p}_i,s_i) \rangle$:
\begin{align}
\label{eq:AFFdef}
\left\langle N({\bm p}_f,s_f) | A_\mu (\bm {q}) | N({\bm p}_i,s_i)\right\rangle  &= 
{\overline u}_N({\bm p}_f,s_f)\left( G_A(q^2) \gamma_\mu
+ q_\mu \frac{\widetilde{G}_P(q^2)}{2 M_N}\right) \gamma_5 u_N({\bm p}_i,s_i) \,, \nonumber \\
\left\langle N({\bm p}_f) | P (\bm{q}) | N({\bm p}_i)\right\rangle  &= 
{\overline u}_N({\bm p}_f) G_P(q^2) \gamma_5 u_N({\bm p}_i) \,, 
\end{align}
where $A_\mu = Z_A \overline{u} \gamma_\mu \gamma_5 d$ and $P = Z_P
\overline{u} \gamma_5 d$ are the renormalized isovector axial and pseudoscalar
currents.  These three form factors have to satisfy, up to
discretization errors, the following PCAC relation, a consequence of the 
axial Ward identity, $\partial_\mu A_\mu - 2 Z_m { m} P=0$:
\begin{equation}
2 {\widehat m} G_P(Q^2) = 2 M_N G_A(Q^2) - \frac{Q^2}{2M_N} {\widetilde G}_P(Q^2) \,,
\label{eq:PCAC}
\end{equation}
where ${\widehat m} \equiv Z_m Z_P (m_u +m_d)/(2 Z_A)$ is the average
bare PCAC mass of the $u$ and $d$ quarks.  Figure~\ref{fig:PCAC},
taken from Ref.~\cite{Park:2021ypf}, shows the degree to which the
axial form factors satisfy Eq.~\protect\eqref{eq:PCAC}, with ME
obtained from 3 analysis strategies: $\{4,3^\ast\}$---standard
analysis with a 3-state fit to the $A_i$ and $P$ 3-point correlators;
$\{4^{N\pi},2^{A_4}\}$---a 2-state fit to the $A_i$ and $P$ 3-point
correlators with $E_1$ determined from a fit to the $A_4$ correlator;
and $\{4^{N\pi},2^{\rm sim}\}$---a simultaneous 2-state fit to all
five $A_\mu$ and $P$ 3-point correlators with $E_1$ left as a free
parameter. The data show that the standard analysis, $\{4,3^\ast\}$,
fails by about 50\% at $M_\pi =
135$~MeV~\cite{Jang:2019vkm}. Including the $N \pi$ state,
($\{4^{N\pi},2^{A_4}\}$ and $\{4^{N\pi},2^{\rm sim}\}$ strategies),
significantly improves the agreement with PCAC, with the remaining
difference attributable to possible discretization errors and/or
contributions of additional excited states. The increase in deviation
(and difference from ``$N \pi$'') as $Q^2 \to 0$ and $M_\pi \to
135$~MeV is correlated with the growth in the difference in $E_1$ between
the ``standard'' and ``$N \pi$'' analyses. $\chi$PT analysis cements
this understanding of enhancement~\cite{Bar:2018xyi}--the axial
current couples to a light pion, and this interaction vertex can be
anywhere in the spatial 3-volume: a well-known observation enshrined
in the pion-pole dominance hypothesis. This volume enhancement causes
a large $N \pi$ contribution.  Thus the case for including the $N \pi$
state in analysis of the axial FF is clear. The question therefore is--what
ES make significant contributions in a given  $C_\mathcal{O}^{3\text{pt}}(\tau;t)$, and including
them.\looseness-1

Knowing that ``$N\pi$'' states contribute to axial/pseudoscalar FF, the precision of
calculations of the axial charge
$g_A$~\cite{Aoki:2019cca,Aoki:2021kgd} is also in question. Again the
issue is the size of the integrated contribution from the tower of $N
(\bm k) \pi(-\bm k)$ excited states.  One way forward, advocated
in~\cite{Park:2021ypf}, is to demonstrate the necessary consistency
check--the value of $g_A$ in the continuum limit from the forward
matrix element has to agree with that obtained by extrapolating the
axial form factor $G_A(Q^2)$ to $Q^2 =0$. This, of course, requires
controlling all the systematics in both calculations.

\section{Electric and magnetic form factors}

The Sachs electric, $G_{E}(Q^2)$, and magnetic, $G_{M}(Q^2)$,
form-factors have been measured extensively in electron-nucleon
scattering experiments.  For the current status of the possible
resolution of the discrepancy in the proton charge radius between the
electron scattering and muon capture experiments see
Ref.~\cite{Xiong:2019umf}. Compared to lattice calculations, the
experimental data for the form factors is very precise and, therefore,
they provide a test for the lattice methodology.

Chiral PT analysis by B\"ar in Ref.~\cite{Bar:2021crj} indicates a
$\approx 5\%$ effect due to the pion loop. The $Q^2$ dependence and
the magnitude of the effect is consistent with the pattern seen in a
summary of world lattice data shown in fig.~[22] in
Ref.~\cite{Jang:2019jkn}. In this case, one expects contributions from vector
current coupling to rho to two pions (vector meson dominance),
however, since the $\rho$-meson is heavy, the enhanced coupling of the
vector current to the $\rho$-meson may be negated by its heavier mass.\looseness-1

\begin{figure*}[tbp] 
{
    \includegraphics[width=0.32\linewidth]{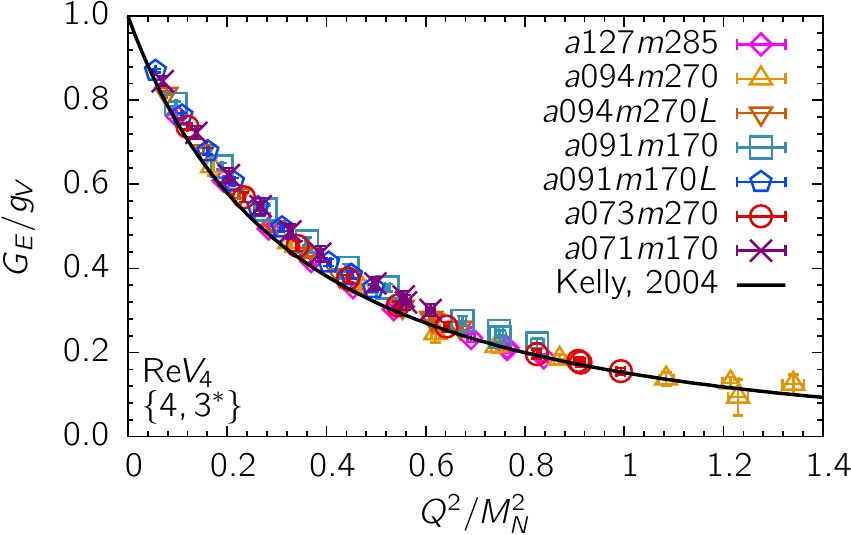}  
    \includegraphics[width=0.32\linewidth]{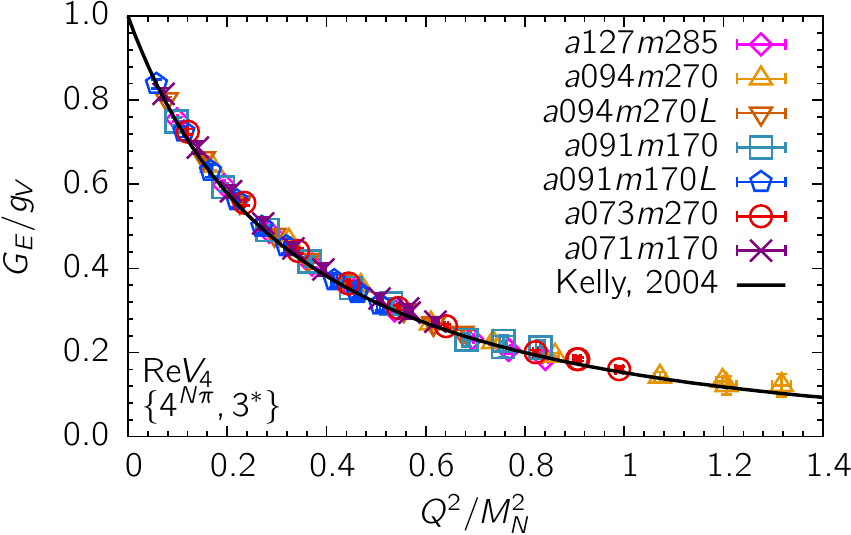} 
    \includegraphics[width=0.32\linewidth]{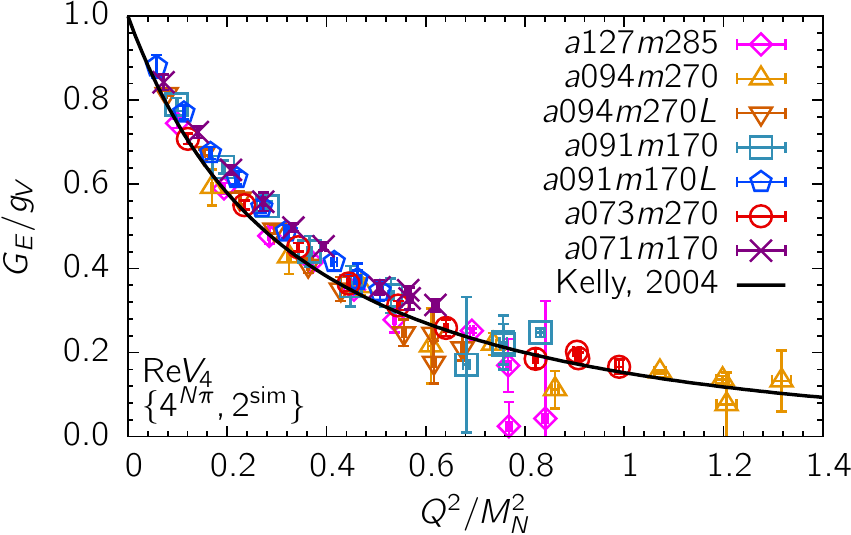}   
}
{
    \includegraphics[width=0.32\linewidth]{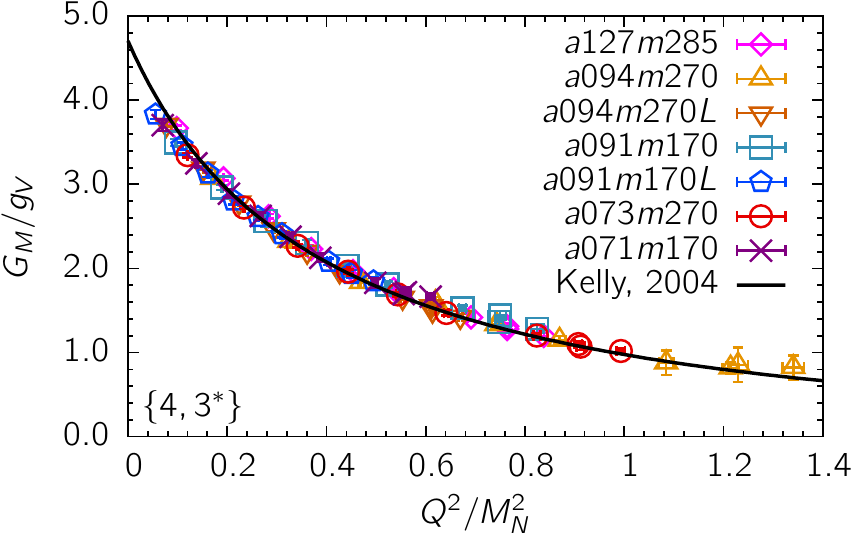}  
    \includegraphics[width=0.32\linewidth]{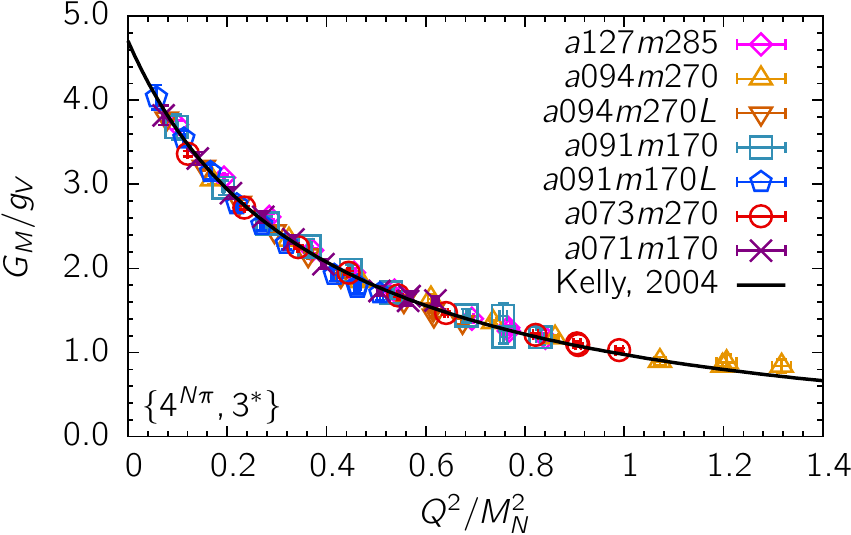} 
    \includegraphics[width=0.32\linewidth]{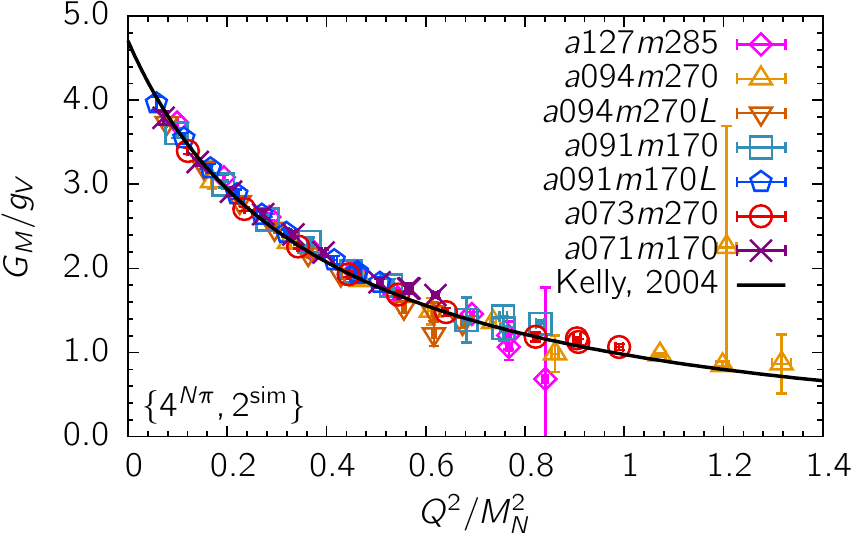}   
}
\caption{$G_E(Q^2)$ and $G_M(Q^2)$ plotted versus
  $Q^2/M_N^2$. Comparison of data extracted using 3 strategies for
  controlling ESC: left labeled $\{4,3^\ast\}$ is the standard 3-state
  analysis; middle labeled $\{4^{N\pi},3^\ast\}$ inputs a narrow prior
  for $E_1$ corresponding to a $N\pi\pi$ state in a 3-state fit; and
  right labeled $\{4^{N\pi},2^{\rm sim}\}$ is a simultaneous 2-state fit
  to all $V_\mu$ channels with $E_1$ left a free parameter. See
  Ref.~\cite{Park:2021ypf} for details.
  \label{fig:GEGMsummary}}
\end{figure*}

Figure~\ref{fig:GEGMsummary} shows our latest data from $2+1$-flavor
clover simulations~\cite{Park:2021ypf} plotted versus $Q^2/M_N^2$. We
find a much better agreement with the Kelly parameterization over the
whole range $0.04 < Q^2 < 1.2$~GeV${}^2$ compared to previous lattice
data~\cite{Jang:2019jkn}.  The differences in form factors between
analyses without (left panel) and with (middle panel) a low-mass
excited state ($N \pi$ or $N \pi \pi$) or with the mass gap determined
from the 3-point functions themselves (right panel) are small. Furthermore, we
observe insensitivity of the data to lattice spacing $a$ and the pion
mass. We are testing this favorable situation, ie, EM form factors showing small systematics, 
by increasing the statistics and adding more ensembles. Once validated, 
lattice QCD is poised to provide precision results in the near
future.\looseness-1

\section{Contribution of the $\Theta$-term to neutron EDM}

The $\Theta$-term $\big( \Theta \frac{i G^a_{\mu\nu} {\widetilde
    G^a_{\mu\nu}} }{32\pi^2}\big)$ is a $P$ and $T$ violating
dimension four operator (also \CPV\ if $CPT$ is conserved) allowed in
the standard model. It contributes an amount $d_n = X \overline\Theta$
to the neutron electric dipole moment (nEDM). Here $\overline\Theta$
is the convention independent coupling and $X \equiv \lim_{q^2\to0}
\frac{F_3(q^2)}{2 M_N \overline\Theta}$ is given by the \CPV\ part of
the electromagnetic form factor $F_3$ calculated using lattice
QCD~\cite{Bhattacharya:2021lol}. Using the current upper bound on the
nEDM, $|d_n| \allowbreak < \allowbreak 1.8 \times 10^{-26}$~$e$~cm
(90\% CL)~\cite{nEDM:2020crw}, on gets ${\overline\Theta} < 10^{-10}$,
an unnaturally small number. Since each \CPV\ interaction contributes
to the nEDM, a value or bound on $d_n$ constraints the parameter space
of possible \CPV\ couplings.  Our goal is to determine the ME that
connect $d_n$ to the couplings, i.e., the analogue of $X$ from $F_3$ in the
$\Theta$-term example, for the Weinberg, quark chromo EDM and 4-fermion
low-energy effective operators up to mass dimension 6 that encapsulate
\CPV\ in the quark-gluon sector~\cite{Pospelov:2005pr} (see talk by
T. Bhattacharya, {\it ibid}).

\begin{figure}[tbp]
    \centering
  \subfigure{
    \includegraphics[width=0.34\linewidth]{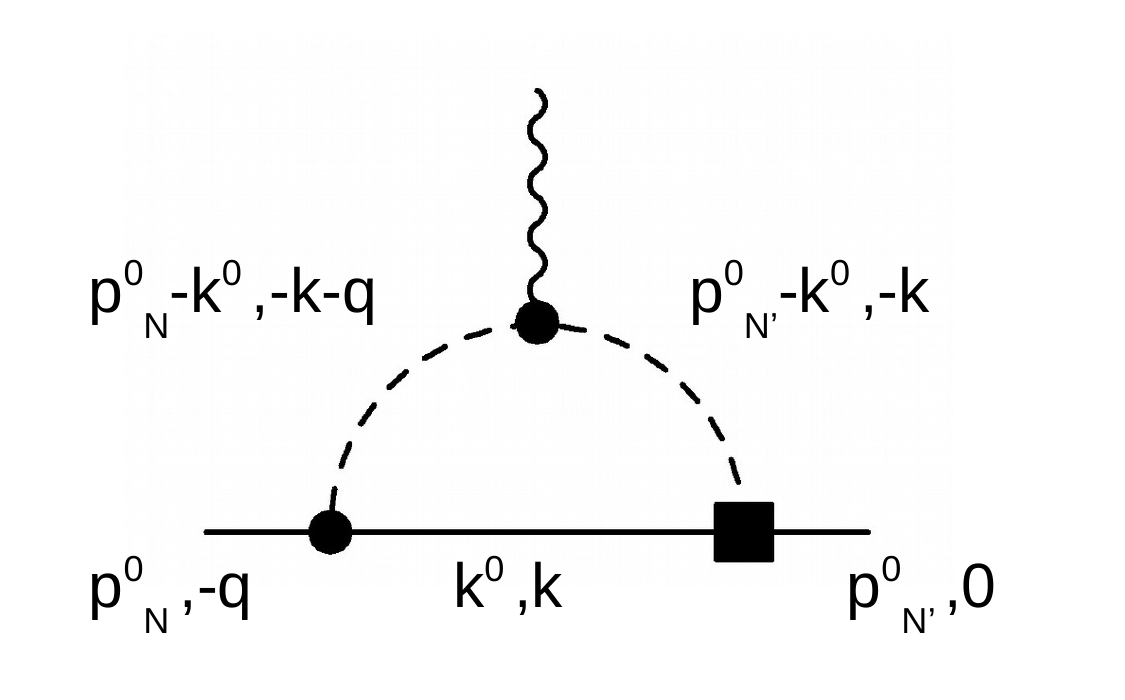}
    \includegraphics[width=0.32\linewidth]{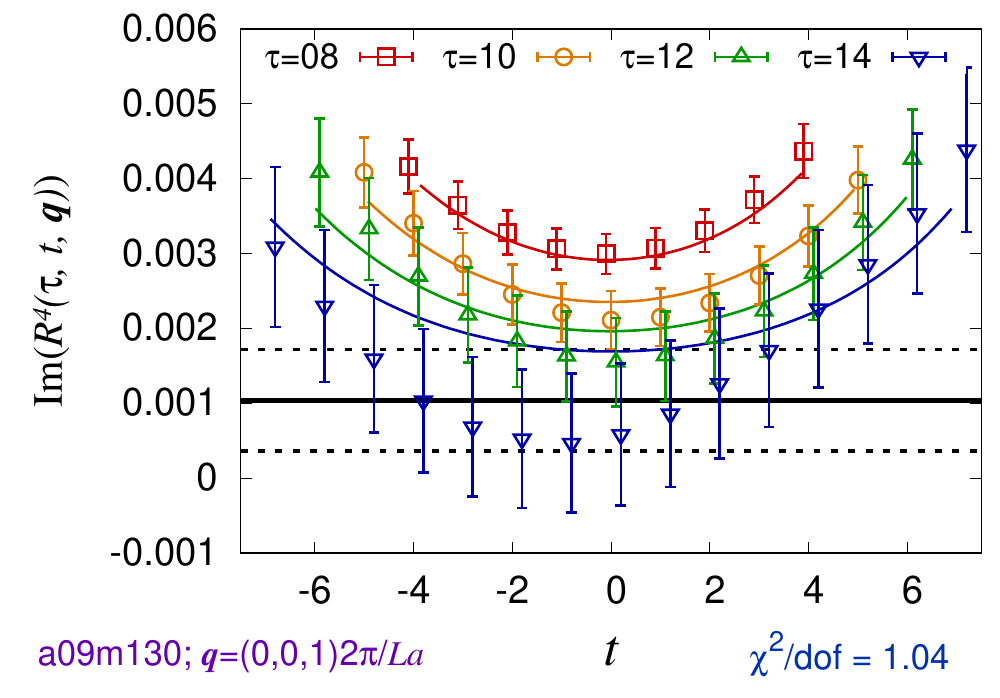} 
    \includegraphics[width=0.32\linewidth]{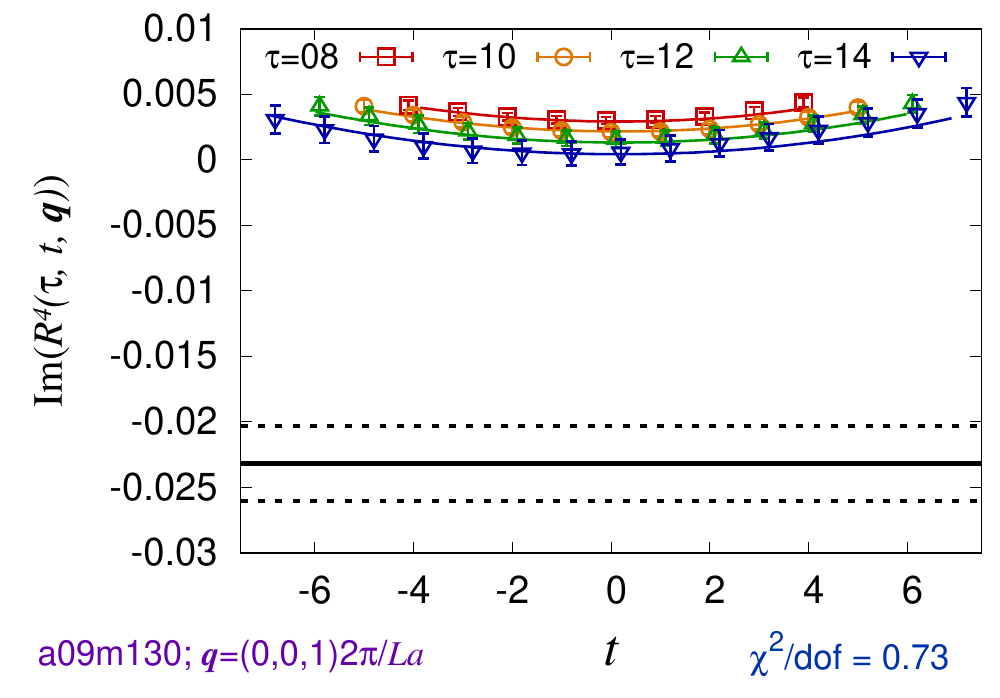} 
}
    \caption{(Left) Leading order diagram for the excited states
      contribution to the three-point function ${\cal C}_{3pt}^\mu$ in
      chiral perturbation theory. A black square denotes an insertion
      of the CP-odd pion-nucleon couplings $\bar g_0$. Filled circles
      denote CP-even pion-nucleon and pion-photon couplings.  A 
      ground state estimate from a two-state fit with $E_1$
      taken from the standard  three-state fit to the two-point function (left
      panel) is compared to that with $E_1$ set equal to the
      non-interacting energy of the $N \pi$ state (right panel).  The
      data are from a $a \approx 0.09$~fm, $M_\pi = 135$~MeV ensemble. The
      $\chi^2$/dof of the two sets of fits are comparable, but the
      extrapolated ground state value (solid black line) is vastly different. 
      All three panels are reproduced from~\protect\cite{Bhattacharya:2021lol}.  
}
    \label{fig:pionloop}
\end{figure}

A $\chi$PT analysis in~\cite{Bhattacharya:2021lol} showed that the gap
between the ground and excited states contributing to the CP-odd
components of the three-point function with the insertion of the
$\bar\Theta$-term is of the order of the pion mass $M_\pi$. Again,
this can be intuitively understood as coming from a long-range pion
loop \cite{Crewther:1979pi} (Fig.~\ref{fig:pionloop} left).  In
Fig.~\ref{fig:pionloop} we also show an example of the big difference
in results (solid black lines) for the ground state matrix elements
extracted from a standard analysis, a three-state fit with $E_1$ from
the two-point function, (middle panel) versus using the
non-interacting energy of the $N \pi$ state (right panel).  The
resulting values of $d_n$ are very different and results including $N \pi$ states are much larger!  Our
work~\cite{Bhattacharya:2021lol} made 2 points: (i) the errors in all
existing lattice calculations of the contribution of the
$\Theta$--term to nEDM are too large to quote a reliable value, and
(ii) resolving the excited state spectrum is essential for precision
determination of $X$; if $N\pi$ states do give dominant contamination, then $d_n$ will be larger and calculable sooner!
An exception to the poor signal in nEDM derived from calculations of the $F_3$ form 
factor ($\Theta$, Weinberg, quark chromo; 3 of 5 effective $D\le
6$ \CPV\ operators arising from BSM) 
is the (fourth) quark EDM operator  that has been determined with $\lesssim 5\%$
uncertainty~\cite{Bhattacharya:2015esa,Gupta:2021ahb}. For it, $X =
g_T^{u,d,s,c,b}$. These tensor charges have
small ESC and are insensitive to the details of the excited-state
spectrum used, ie, whether $N\pi$ states are included in the fits to remove ESC~\cite{Aoki:2019cca,Aoki:2021kgd,Bhattacharya:2015esa}.  \looseness-1

\vspace{-0.2in}

\section{The pion-nucleon sigma term}

The pion--nucleon $\sigma$-term is $\spiN \equiv { m}_{ud}\, g_S^{u+d}
\equiv {m}_{ud} \, \langle N({\kk},s)| \bar{u} u + \bar{d} d |
N({\kk},s) \rangle$. The scalar charge $g_S^q$ is the forward
matrix element of the scalar density $\bar{q} q$ between the nucleon
ground state, 
\begin{align}
g_S^q =  \langle N({\kk}=0,s)| \bar{q}  q | N({\kk}=0,s) \rangle.
\label{eq:gSdef}
\end{align}

The $\spiN$ is a fundamental parameter of QCD--it quantifies the
amount of the nucleon mass that comes from $u$- and $d$-quark masses
being non-zero. Also, the scalar charge $g_S$ enters into
the spin independent cross-section of dark matter with nuclear
targets~\cite{Bottino:2001dj,Hoferichter:2017olk}, lepton flavor violation in $\mu\to e$ conversion in
nuclei~\cite{Cirigliano:2009bz,Crivellin:2014cta}, and in electric
dipole
 moments~\cite{Engel:2013lsa,deVries:2016jox}. Thus
knowing $\spiN$ and $g_S^q$ accurately is important. In addition to lattice
calculations, $\spiN$ has also been extracted phenomenologically from $\pi-N$ scattering via
the Cheng--Dashen low-energy theorem~\cite{Cheng:1970mx,Brown:1971pn}.

The current status of lattice calculations of $\spiN$ and comparison
to phenomenology has been reviewed by
FLAG~\cite{Aoki:2019cca,Aoki:2021kgd}. The reviewed results show a tension between
the lattice estimates that favor $\spiN \approx 40$~MeV versus
values from phenomenology centered around $\spiN \approx 60$~MeV~\cite{Hoferichter:2015dsa,RuizdeElvira:2017stg}.

\begin{figure}[tbp] 
\vspace{-0.1in}
    \centering
  \subfigure{
    \includegraphics[width=0.40\linewidth]{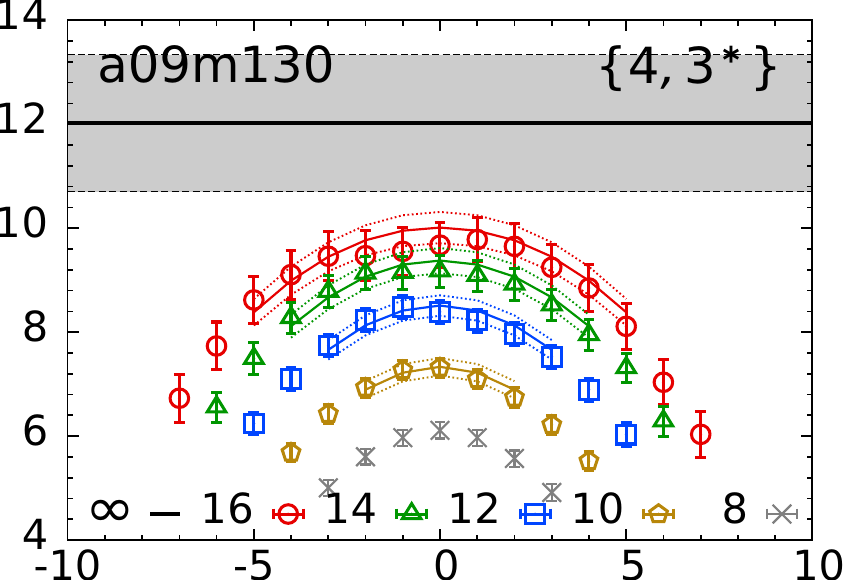} \hspace{0.5cm}
    \includegraphics[width=0.40\linewidth]{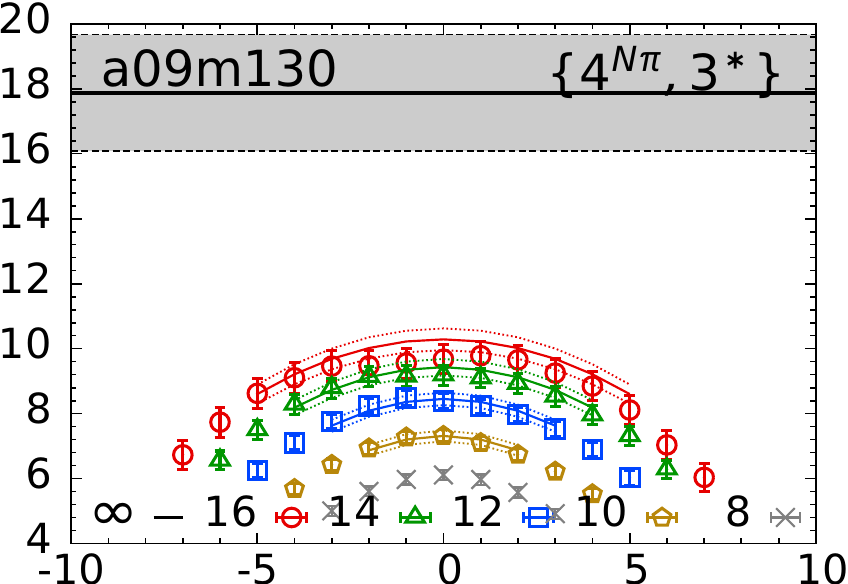} 
}
    \caption{The data for $g_S^{u+d+2l}$ from the physical mass,
      $a09m130$ ensemble and the two fits to remove ESC.  (Left)
      standard analysis and (Right) including $N \pi$ and $N \pi \pi $
      states.  Both panels are reproduced
      from~\protect\cite{Gupta:2021ahb}.  }
    \label{fig:sigma}
\end{figure}

Our recent lattice calculation~\cite{Gupta:2021ahb} has been performed in the
isospin symmetric limit, i.e., with ${m}_{ud} = (m_u + m_d)/2$ the
average of the light quark masses. The N${}^2$LO $\chi$PT analysis showed that 
there is an enhanced contribution from $N \pi$ and $N \pi \pi $ states due to 
the large coupling of the scalar source to two pions, i.e., a large quark condensate. 
Our result with the standard analysis is $\spiN \approx 40$~MeV (consistent with 
previous lattice estimates) while the one including contributions of $N \pi$ and $N \pi \pi $ states 
gave $\spiN \approx 60$~MeV, which is consistent with phenomenology. 
The data from the physical mass, $a09m130$, ensemble and the two fits
to remove ESC are shown in Fig.~\ref{fig:sigma}. Again, the two fits
with very different results are not differentiated by the $\chi^2$. To
reach discrimination will, we estimate, require similar precision data
at $\tau = 18$ and 20, i.e., a $\ge $10X increase in
statistics.\looseness-1

\section{Conclusions}

The lattice methodology for the calculations of nucleon matrix
elements and their extraction from correlation functions using spectral 
decomposition is robust, however, the two related issues of the exponential
decay of the statistical signal with source-sink separation in all
nucleon n-point functions and the contribution of low-lying
multihadron excited states have to be addressed before precision
results can be obtained. Here, we review three examples, (i) the
axial form factors $ G_A(q^2), {\widetilde{G}_P(q^2)}\ {\rm
  and}\ G_P(q^2)$; (ii) the $\Theta$-term contribution to nEDM; and
(iii) the pion-nucleon sigma term, for which $\chi$PT indicates a
large contribution and including ``$N\pi$'' changes the results very
significantly. The case for including $N \pi$ states in the analysis of 
axial form factors is strengthened by (i) the strong $\chi^2$  preference  
in fits to the 3-point function with the insertion of the $A_4$
operator and (ii) the
resulting form factors satisfy the PCAC relation to within expected
discretization errors. Unfortunately, for all other quantities, fits
to the current lattice data do not distinguish between the
``standard'' and ``$N\pi$'' analyses on the basis of $\chi^2$. (The
weakness in the ``$N \pi$'' analysis for most quantities is the 
narrow priors used for excited state energies.) Thus, we 
resort to $\chi$PT for guidance on which analysis to choose. In
future, with increased statistics, we anticipate fits can be made
directly to 3-point functions with one or even two excited states 
and their energies $E_i$ left as free parameters.\looseness-1 

Another point in support of taking input from $\chi$PT in the ESC fits
is that results of the same analysis are used for making the chiral fits to
quantify the behavior versus $M_\pi^2$ and get the final results at
the physical pion mass and in the continuum limit. For self
consistency, one should include the states that make a significant
contribution in both parts of the analysis.

Promising future approaches include (i) to overcome the signal to noise
problem develop methods 
based on analytic continuation of the contour of integration 
(see Refs.~\cite{Wagman:2016bam,Detmold:2021ulb}), and (ii) to 
use a variational basis of nucleon interpolating operators that includes 
multihadron operators to project on to the ground state of the nucleon
at much earlier source-sink separations. Hopefully, one or more of
such novel methods will, in the near future, break the logjam.

\acknowledgments
The calculations presented are based on two sets of ensembles: the
$2+1+1$-flavor HISQ ensembles generated by the MILC collaboration and
the $2+1$-flavor Wilson-clover ensembles generated by the
JLAB/W\&M/LANL/MIT collaborations. The calculations used the Chroma
software suite~\cite{Edwards:2004sx}. We gratefully acknowledge
computing resources provided by NERSC, OLCF at Oak Ridge, USQCD and
LANL Institutional Computing. Support for this work was provided the
U.S. DOE Office of Science HEP and NP, the SNSF, and by the LANL LDRD program.

\bibliographystyle{JHEP}
\let\origbibitem\bibitem
\def\bibitem#1#2\emph#3, {\origbibitem{#1}#2}
\bibliography{ref} 


\end{document}